\documentclass[10pt,aps,prx,twocolumn,nofootinbib,superscriptaddress,longbibliography]{revtex4-2}

\makeatletter\let\switch@array\relax\makeatother

\usepackage{amsmath,amsfonts,amssymb,bm,bbm}
\usepackage{graphicx}
\usepackage[colorlinks=true, linkcolor=blue, citecolor=blue, urlcolor=blue, breaklinks=true]{hyperref}
\usepackage{physics}
\usepackage{dsfont}
\usepackage{bbold}
\usepackage{soul}
\usepackage{enumitem}
\usepackage[table]{xcolor}
\usepackage{tikz}
\usetikzlibrary{arrows.meta, positioning}

\definecolor{purple}{HTML}{9326ff}

\newcommand{\figpanel}[1]{\textbf{\textsf{#1}}}
\newcommand{\pdagger}{{\phantom{\dagger}}}

\definecolor{light_blue}{HTML}{f0f5ff}
\definecolor{light_grey}{HTML}{ededed}
\definecolor{check}{HTML}{ff7300}
\definecolor{energy_orange}{HTML}{ffb061}

\newcommand{\methods}{\hyperlink{methods}{Methods}}
\newcommand{\data}{\hyperlink{data}{Data Availability}}

\usepackage{tabularx}
\newcolumntype{Y}{>{\centering\arraybackslash}X}
\usepackage{tcolorbox}
\definecolor{light_gray}{HTML}{f2f2f2}
\tcbset{fontupper=\normalsize,
colback=white, colframe=black, colbacktitle=light_gray,
coltitle= black, arc=0.5mm, center title}

\tcbset{fontupper=\normalsize,
	colback=white, colframe=black, colbacktitle=light_grey,
	coltitle= black, arc=0.25mm}
\usepackage{latexsym,bm,amsfonts,amstext,graphicx,bbm,relsize,dsfont,lipsum,bold-extra,amsmath,amssymb,enumerate,float}
\usepackage[bb=dsserif]{mathalpha}

\usepackage{tcolorbox}
\tcbuselibrary{skins}

\newcommand{\us}{$\mathrm{\mu s}$}
\newcommand{\um}{$\mathrm{\mu m}$}

\usepackage{titlesec}
\titleformat*{\section}%
    {\fontsize{12}{12}\bfseries}%
    {}

\titlespacing*{\section}
  {0pt}         
  {10pt}        
  {10pt}         

\titleformat*{\subsection}%
    {\fontsize{10}{10}\bfseries}%
    {}

\titlespacing*{\subsection}
  {0pt}         
  {10pt}        
  {5pt}         

\titlespacing*{\subsubsection}
  {0pt}         
  {10pt}        
  {5pt}         
  
\usepackage{titlesec}
\titleformat*{\subsubsection}%
    {\fontsize{10}{10}\itshape}%
    {}





\begin{document}

\title{Validation and calibration of quantum hardware\\ through the many-body quantum Mpemba effect}

\author{Francesco Campaioli}
\email{francesco.campaioli@rmit.edu.au}
\affiliation{Department of Physics, School of Science, RMIT University, Melbourne, 3000, Victoria, Australia}
\affiliation{RMIT Applied Quantum Technologies Centre, RMIT University, Melbourne, 3000, Victoria, Australia}
\affiliation{Dipartimento di Fisica e Astronomia, Universit\`a degli Studi di Padova, 35131 Padova, Italy}

\author{Marco Avesani}
\affiliation{Dipartimento di Ingegneria dell’Informazione, Universit\`a degli Studi di Padova, 35131 Padova, Italy}
\affiliation{Padua Quantum Technologies Research Center, Universit\`a degli Studi di Padova, Padova, Italy}
\author{Oren Raz}
\affiliation{Department of physics of complex systems, Weizmann Institute of Science, Rehovot, Israel, 76100}

\author{Roderich Moessner}
\affiliation{Max Planck Institute for the Physics of Complex Systems, N\"othnitzer Str.~38, 01187 Dresden, Germany}

\author{Gianluca Teza}
\email{teza@pks.mpg.de}
\affiliation{Max Planck Institute for the Physics of Complex Systems, N\"othnitzer Str.~38, 01187 Dresden, Germany}
\affiliation{Dipartimento di Fisica, Universit\`a di Trieste, Strada Costiera 11, I-34151 Trieste, Italy}

\date{\today}

\begin{abstract}
We introduce a validation process that harnesses engineered many-body relaxation to control and calibrate quantum hardware.
On two independently developed neutral-atom processors, we realize the many-body quantum Mpemba effect in an open system for the first time.
Initial-state engineering creates fast and slow relaxation pathways: the fast pathway opens access to unreachable mixed-state physics before hardware noise obscures the target dynamics, whereas the slow pathway amplifies hidden imperfections in preparation and control.
Computational-basis measurements directly and independently benchmark dynamical reliability, revealing each processor's actual operating window as a many-body simulator.
The processors are thus judged by the very dynamics they are built to reproduce.
Complementary responses disentangle control errors and drive an adaptive, time-resolved scheme that supplies hardware developers directly actionable corrections, enhancing faithful reproduction of the target dynamics.
These results establish many-body relaxation as a transferable validation and calibration tool for programmable quantum processors.
\end{abstract}
\maketitle
\makeatletter

{\noindent 
Before silicon rose to dominance, early electronic computers relied on bulky, fragile, and power-hungry vacuum tubes~\cite{haigh2021new}.
The invention of the solid-state transistor in the late 1940s sparked a competition between alternative materials such as germanium and silicon~\cite{bardeen1948transistor}, whose outcome was settled only after years of systematic testing, characterization, and industrial development~\cite{Riordan1999}. Today, quantum technologies face a similar stage of rapid diversification, with a wide range of platforms competing to reach utility-scale performance~\cite{ladd2010quantum, deLeon2021, Kim2023}. 
Leading approaches include trapped ions~\cite{brown2021materials, Malinowski2023,fossfeig2025trappedion}, superconducting circuits~\cite{Kjaergaard2020, Bravyi2022, Acharya2025}, photonic~\cite{Slussarenko2019, Wang2020c, maring2024versatile}, and neutral-atom arrays~\cite{Henriet2020quantumcomputing, evered2023high, Radnaev2025}. These platforms offer complementary trade-offs in coherence, connectivity, and control, making their relative performance strongly task-dependent and difficult to assess in a unified manner. However, validation and calibration remain fundamentally challenging: exact techniques such as quantum state and process tomography provide complete information but scale exponentially with system size~\cite{Christandl2012,Tran2021,Caprotti2024}, while scalable alternatives such as randomized benchmarking achieve polynomial cost by sampling over a restricted set of input states or circuits, at the price of simplifying assumptions about the underlying noise~\cite{Knill2008,daSilva2011,Proctor2022}. This challenge is especially acute for quantum simulations, where dedicated validation protocols remain scarce and markedly less mature than their quantum-computing counterparts~\cite{Eisert2020}. Any advance that improves the accuracy of validation while reducing its cost in sequence length and number of measurements is therefore of immediate practical value.

\begin{figure*}[ht!]
    \centering
    \includegraphics[width=\textwidth]{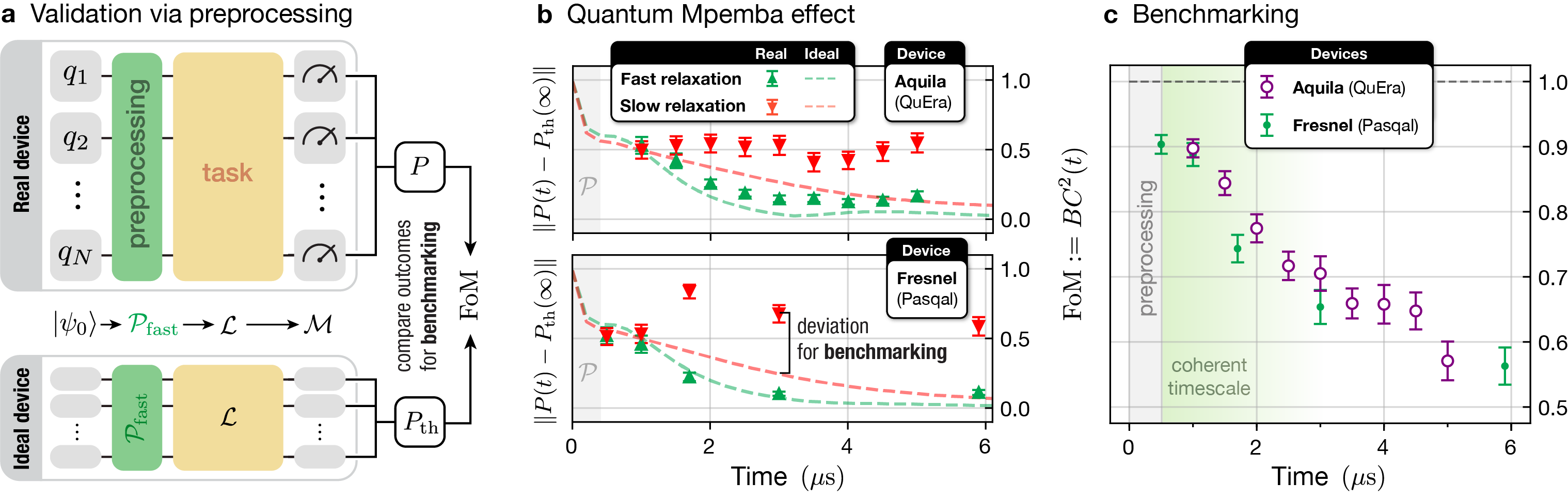}
    \caption{\textbf{Overview of validation routine.} \figpanel{a}, A register of $N$ qubits $q_i$, initialized in a state $\ket{\psi_0}$ on a real device, undergoes a preprocessing sequence designed to accelerate ($\mathcal{P}_\mathrm{fast}$, \textit{green}) or decelerate ($\mathcal{P}_\mathrm{slow}$, \textit{red}) the relaxation dynamics induced by noise for a given task $\mathcal{L}$. After the task is executed, the qubits are measured via a process $\mathcal{M}$. The resulting measurement outcomes $P$ from the real device are compared against simulated outcomes $P_\mathrm{th}$ (ideal device) to benchmark device performance; deviations are attributed to hardware imperfections and systematic errors. \figpanel{b}, Measurement outcomes showing fast and slow relaxation towards the equilibrium configuration, obtained on two neutral-atom devices: Aquila (QuEra, \textit{top}) and Fresnel (Pasqal, \textit{bottom}). The shaded gray band at short times marks the preprocessing window. Green upward triangles (fast) and red downward triangles (slow) indicate the distance from equilibrium for two preprocessing sequences, with dashed lines denoting the corresponding theoretical predictions; error bars represent one standard deviation obtained from bootstrapping. On both devices, fast preprocessing leads to accelerated relaxation despite starting further from equilibrium: a signature of the Mpemba effect. Deviations of the data from theory, shown here before recalibration, quantify the errors introduced by the device and are used to evaluate its quality and detect systematic biases. \figpanel{c}, Device performance is estimated by evaluating a figure of merit ($\mathrm{FoM}$) derived from the difference between measured and theoretical system configurations. Here, we evaluate the \textit{Bhattacharyya coefficient} $BC^2$ of Eq.~\eqref{eq:BC_coefficient} for Aquila (\textit{open purple circles}) and Fresnel (\textit{filled green circles}) as a function of time, where $BC^2 = 1$ corresponds to ideal quality and provides an upper bound on the true quantum fidelity of the devices~\cite{Eisert2020}. Error bars denote one standard deviation obtained via bootstrapping. The green shaded area indicates the time window during which the devices operate predominantly coherently, beyond which decoherence becomes prominent.
}
    \label{fig:graphical_abstract}
\end{figure*}

Here we address this challenge by turning noise, an unavoidable limitation of quantum hardware, into a diagnostic tool.
Qubit registers on current quantum platforms are inherently open many-body systems, whose relaxation dynamics are shaped by collective effects~\cite{Cattaneo2023, PRXQuantum2023, Fazio2025}. The latter enrich the structure of the relaxation landscape, enhancing the sensitivity of the dynamics to initial conditions via metastability~\cite{Macieszczak2021}. A paradigmatic example is the quantum Mpemba effect~\cite{Moroder2024,Ares2025}---a generalization to quantum systems of the classical phenomenon whereby hot water can freeze faster than cold~\cite{Teza2026}---in which selective state preparation can substantially accelerate or slow down convergence to equilibrium~\cite{Beato2026}.
As quantum platforms and many-body theory push each other forward, such nonequilibrium phenomena become experimentally accessible, opening new opportunities for the development of reliable quantum hardware.
In this work we propose a new use for this synergy: a hardware validation principle that harnesses the many-body quantum Mpemba effect.
By preparing distinct initial states and tracking their relaxation trajectories, we gain direct access to the extremal dynamical modes governing the device, corresponding to the fastest and slowest relaxation processes. As we show, this enables a stringent benchmark of device performance and a direct route to hardware calibration that exploit many-body effects to go beyond the reach of conventional single- and two-qubit diagnostics.

To put this principle into practice, we implement a Mpemba-based protocol for the validation and calibration of quantum simulators. We demonstrate this protocol on two independent neutral-atom devices, QuEra’s Aquila~\cite{wurtz2023aquila} and Pasqal’s Fresnel~\cite{Silverio2022pulseropensource}, implementing a dissipative six-atom Ising ring and selectively accelerating or slowing down its relaxation to equilibrium. The use of two independent platforms shows that our protocol provides a reproducible validation primitive that transfers across hardware, while revealing both robust shared features and complementary device-specific imperfections. The resulting out-of-equilibrium dynamics provide a sensitive probe of imperfections in state preparation and control sequences. From computational-basis measurements alone, we extract an upper bound on the quantum fidelity as a figure of merit and directly compare the performance of the two devices. The same data then yield a calibration procedure that returns concrete corrections to the implemented pulse sequences.

By targeting convergence to equilibrium, central in optimization, our approach evaluates device performance on a specific task. This setting is closer to end-user applications than general device-level metrics like coherence times, in analogy with modern CPU and GPU benchmarks. Altogether, this framework establishes an efficient and platform-independent route to benchmarking and calibration of quantum devices, built to leverage nonequilibrium dynamics.

\subsection*{Quantum validation through Mpemba effect}
\noindent
The validation primitive that we propose in this work consists in inserting a \textit{preprocessing sequence} before running a task on the quantum device. The preprocessing is constructed to control sensitivity to noise, by accelerating or slowing-down the relaxation of the qubits register. For the sake of simplicity, let us illustrate the idea by considering a quantum simulation task, schematically represented in Fig.~\ref{fig:graphical_abstract}~\figpanel{a}.
For example, let us consider a device initialized in a fiducial state $\ket{\psi_0}$, and a task represented by a sequence $\mathcal{L}$, uniquely associated to some time-independent Hamiltonian $H$ and noise model $\mathcal{D}$, followed by some measurement\footnote{The measurement process can be most generally described by a Positive Operator-Valued Measure (POVM)~\cite{nielsen_chuang_2000, renes_etal_2004}.} $\mathcal{M}$. The device performance is then determined by evaluating some figure of merit $\mathrm{FoM}\in[0,1]$ on the measurement outcomes, e.g. the quantum state fidelity~\cite{Eisert2020}, expressing the quality of the sequence implementation such that $\mathrm{FoM}=1$ for ideal accuracy.  

In this settings, we insert a \textit{preprocessing sequence} $\mathcal{P}$ before running the task. Importantly, we consider a pair of preprocessing sequences, $\mathcal{P}_\mathrm{fast}$ and $\mathcal{P}_\mathrm{slow}$, chosen to prepare the system in states that are, respectively, rapidly and slowly affected by noise. As we discuss in the next section, these preprocessing are designed to minimize and maximize the overlap of the preprocessed state $\mathcal{P}\ket{\psi_0}$ with the slowest-decaying modes of the dynamics $\mathcal{L}$ and, therefore, they depend on the task sequence and the noise model of the device. Importantly, these preprocessing sequences induce a Mpemba effect for the Liouville superoperator $\mathcal{L}$~\cite{Teza2026}. 
This choice opens to evaluating the performance of the device on the states that are \textit{most} and \textit{least} affected by noise during the task, within the constraints imposed by the device. In many-body systems, this preprocessing can induce high sensitivity to errors in the fiducial state and pulse sequences for implementing the task $\mathcal{L}$, since the conditions for fast and slow relaxation occupy a narrow region of parameter space, as shown in Fig.~\ref{fig:spectrum_and_optimisation} in \methods{}.

Before discussing our demonstration of this approach, let us make a few remarks. 
First, our approach can be generalized to time-dependent Hamiltonians $H(t)$---thus covering a rather broad class of tasks in gate-based and analog devices---using a variety of tools, like Floquet engineering~\cite{Mori2023} and Magnus expansion~\cite{BLANES2009151}.
Second, let us note that the preprocessing sequences do not alter the task itself, and are instead used to evaluate the performance of the device for the considered task via the chosen $\mathrm{FoM}$, as shown in Fig.~\ref{fig:graphical_abstract}~\figpanel{c}. Third, this approach can also be used to calibrate devices, by minimizing the loss function $\varepsilon :=1 - \mathrm{FoM}$ over the space of control sequences, offering a cost-effective on-chip calibration routine and a powerful prescriptive tool to improve devices. 

Let us also note that our approach is compatible with any quality measure $\mathrm{FoM}$, the choice of which should reflect both the capabilities of the device and the task of interest. For instance, tasks like quantum annealing~\cite{kadowaki1998quantum} and variational quantum eigensolvers (VQEs)~\cite{peruzzo2014variational} can be validated based on outcomes in the computational basis alone. Other tasks, by contrast, require informationally complete figures of merit, such as the quantum fidelity, which become prohibitively expensive to evaluate as the register size $N$ grows. Crucially, the figures of merit we consider here exploit many-body properties of the system, going beyond single- and two-qubit diagnostics. A variety of many-body figures of merit are available for this purpose, ranging from well-established measures such as the trace distance to more recent proposals such as the \textit{quantum many-body score}~\cite{erbin2026many}. 

Here we adopt an informationally accessible $\mathrm{FoM}$ based on the \textit{Bhattacharyya coefficient} ($BC$), which exploits all the information available from measurements in the computational basis. As discussed later, we use the $BC$ to upper-bound the true quantum fidelity of the devices under consideration~\cite{Watrous_2018}. While a lower bound would be desirable, this is not possible using an informationally incomplete measurement scheme based on the computational basis only~\cite{guhne2007toolbox, seevinck2008partial, Flammia2011}.
\begin{figure*}
    \centering
    \includegraphics[width=0.95\textwidth]{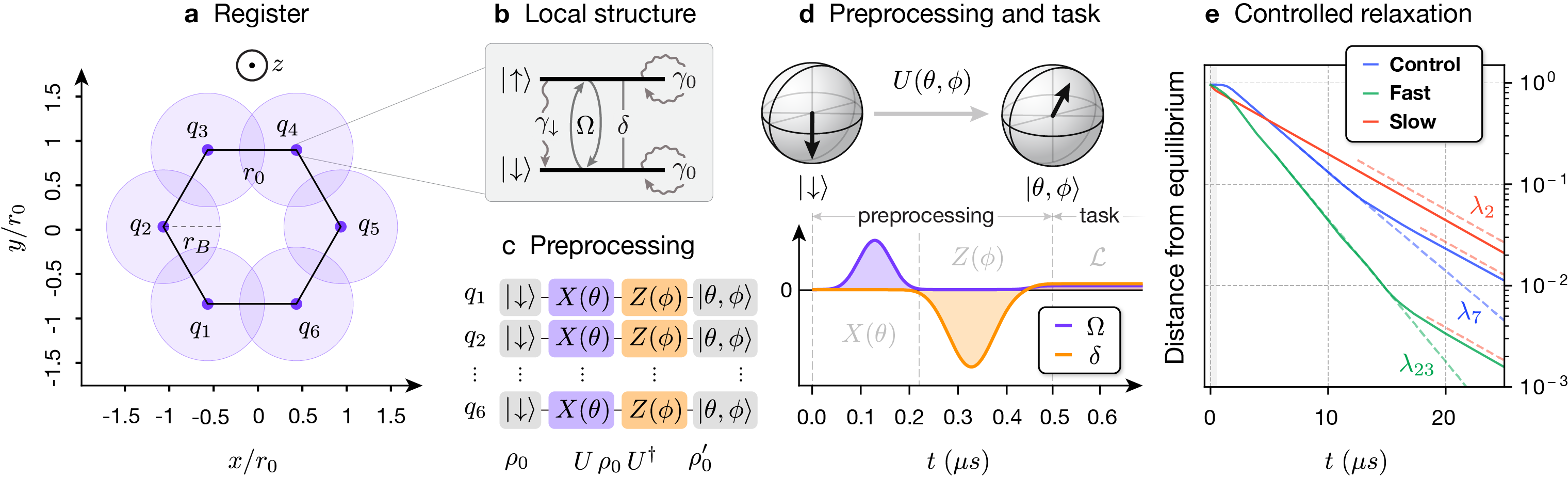}
    \caption{\textbf{Implementation of validation routine on Rydberg atom devices.}  \figpanel{a}, We program a registers of six Rydberg atoms, arranged in a reproducible and exactly-solvable hexagonal layout with lattice spacing $r_0$, forming a ring of 6 qubits ($q_i$) on the $xy$-plane of the device trap. Ising interactions between atoms pairs $(i,j)$ depend on distance $r_{ij}^{-6}$~\cite{wurtz2023aquila,Silverio2022pulseropensource}, as in Eq.~\eqref{eq:SimulatorHamiltonian}. The shaded area around the atoms represents the blockade radius $r_B$, which is a measure of the interaction range. \figpanel{b}, Atoms are controlled by programmable transverse and longitudinal fields via globally addressed coupling strength $\Omega$ and $\delta$, respectively, as shown in Eq.~\eqref{eq:SimulatorHamiltonian}. The atoms undergo device-dependent local relaxation from the Rydberg state $\ket{\uparrow}$ to the ground $\ket{\downarrow}$ state at rate $\gamma_\downarrow$, and dephasing on the ground-Rydberg basis $\mathcal{B}:=\{\ket{\downarrow},\ket{\uparrow}\}$ at rate $\gamma_0$. \figpanel{c}, To control the speed of relaxation, we apply a preprocessing sequence that approximates the unitary operation $U(\theta,\phi) = \otimes_{i=1}^{6} Z_i(\phi)X_i(\theta)$, where $Z_i(\phi)$ and $X_i(\theta)$, defined in Eqs.~\eqref{eq:Z} and~\eqref{eq:X}, are rotation on the Bloch sphere of qubit $i$ around the $z$ and $x$ axes, respectively. This maps the initial fiducial state of the system $\rho_0 = \ketbra{\downarrow \downarrow \cdots \downarrow}{\downarrow \downarrow \cdots \downarrow}$ to a preprocessed state $\rho'_0$, which is approximately separable, before executing the simulation task $\mathcal{L}$. \figpanel{e}, Illustration of fast (slow) relaxation, achieved by optimizing the preprocessing angles $(\theta,\phi)$ to minimize (maximize) the overlap between the preprocessed state $\rho'_0$ and the slowest decaying mode $L_2$ of the open system dynamics, associated with decay rate $\lambda_2$. Fast relaxation can display an initial decay rate that is significantly larger ($\lambda_{23}$) than both $\lambda_2$ and the decay rate of the control state ($\lambda_7$), obtained for $U(0,0)=\mathbbm{1}$.}
    \label{fig:overview}
\end{figure*}

\subsection*{Preprocessing on Rydberg atom devices}
\noindent
Let us now discuss how we obtain the fast and slow preprocessing sequences for an analog simulation task on QuEra's Aquila and Pasqal's Fresnel Rydberg atom quantum simulators. These quantum devices operate by programming time-dependent sequences in registers of Rubidium atoms trapped on a plane using optical tweezers and controlled using laser fields, as illustrated in Fig.~\ref{fig:overview} \figpanel{a}. The atoms' dynamics is governed by the ground-Rydberg Hamiltonian
\begin{equation}
    \label{eq:SimulatorHamiltonian}
    H_\mathrm{gR} = \frac{\Omega_t}
    {2}\sum_{i=1}^N\sigma_i^{(x)}-\delta_t\sum_{i=1}^N n_i+\sum_{i<j}\frac{C_6}{r_{ij}^6} n_i n_j ,
\end{equation}
where where $n_i:= (\mathbbm{1}_i+\sigma_i^{(z)})/2$ is the local excitation operator, $\sigma_i^{(x)}, \; \sigma_i^{(y)}, \; \sigma_i^{(z)}$, are Pauli operators acting on the $i$-th atom,  $\Omega_t$ is the programmable transverse field strength, known as \textit{amplitude}, $\delta_t$ is the programmable longitudinal field strength, known as \textit{detuning},  and $r_{ij}$ is the inter-atom distance. The coupling strength between atoms depends on the device-specific interaction constant $C_6$, which is provided in Table~\ref{tab:device_params}. 
The atoms also undergo local relaxation and dephasing on the ground-Rybderg basis $\mathcal{B}:=\{\ket{\downarrow}, \ket{\uparrow}\}$ (i.e, the $z$-basis), due to their interaction with the environment, as shown in Fig.~\ref{fig:overview} \figpanel{b}. These transitions are represented by local transition operators
\begin{align}
    \label{eq:relax}
    & J^{\downarrow}_i = \sqrt{\gamma_\downarrow} \sigma_i^{(-)} = \sqrt{\gamma_\downarrow} \frac{\sigma_i^{(x)}-\sigma_i^{(y)}}{2}, \\
    \label{eq:deph}
    & J^{0}_i = \sqrt{\gamma_0} \sigma_i^{(z)},
\end{align}
with $\gamma_\downarrow$ and $\gamma_0$ being device-specific relaxation and dephasing rates, respectively, in the 10-100 kilohertz range (See Tab~\ref{tab:device_params}). 

Here, we program an atom register given by an hexagonal ring with lattice constant $r_0$, shown in Fig.~\ref{fig:overview}~\figpanel{a}, and simulate the dynamics of a target transverse-field quantum Ising model $H = \Gamma \sum_{i=1}^N \sigma_i^{(x)} + \sum_{i<j}^N V_{ij}^\pdagger \sigma_i^{(z)}\sigma_i^{(z)}$ by setting $(\Omega, \delta, C_6/r_0^6) \to (2\Gamma, 4V, 4V)$~\cite{Kochsiek2022}. The choice of system size and geometry is guided by reproducibility and exact solvability, necessary for validating proof-of-concept demonstrations and benchmarking tasks. The relaxation dynamics of the register is governed by the Gorini-Kossakowski-Sudarshan-Lindblad (GKSL) Markovian quantum master equation $d\rho_t/dt =: \dot{\rho}_t =\mathcal{L}[\rho_t]$, where the Liouville superoperator generating the dynamics is given by
\begin{equation}
    \label{eq:lindblad}
    \mathcal{L}[\rho_t], \\
    = -i[H,\rho_t]+\sum_l \left( J_l^\pdagger\rho_t J_l^\dagger-\frac{1}{2}\left\{J_l^\dagger J_l^\pdagger,\rho_t\right\}\right),
\end{equation}
with $\rho_t$ being the density operator representing the state of the system at time $t$, and $J_l$ are quantum jump operators of Eqs.~\eqref{eq:relax},~\eqref{eq:deph}~\cite{Breuer2002}, with $\hbar\equiv 1$. This open system model has a unique equilibrium state $\rho_{(\infty)}$ as long as $\Gamma, V, \gamma_0, \gamma_\downarrow \neq 0$.

To obtain the preprocessing sequence, we optimized over rotations that act identically on each qubit\footnote{This choice is imposed by the global addressing constraint on Pasqal's machine Fresnel.}, by approximating the unitary $U(\theta,\phi)=\bigotimes_{i=1}^N Z_i(\phi)X_i(\theta)$, where
\begin{align}
    \label{eq:Z}
    &Z_i(\phi) := \exp\left(+i\phi\sigma_i^{(z)}/2\right), \\
    \label{eq:X}
    &X_i(\theta) := \exp\left(-i\theta\sigma_i^{(x)}/2\right).
\end{align}
This sequence maps the initial fiducial state $\ket{\downarrow}_i$ of each atom onto an arbitrary local state $\ket{\theta,\phi}_i$, which can be represented as a vector $v(\theta,\phi)$ on the Bloch sphere (see Fig.~\ref{fig:overview}~\figpanel{d} and details in \methods{}). To accelerate (slow-down) the approach to equilibrium, we optimize $\theta$ and $\phi$ by minimizing (maximizing) the overlap
between the preprocessed state $\rho'_0 := U(\theta,\phi)\ketbra{\downarrow \downarrow \cdots \downarrow}{\downarrow \downarrow \cdots \downarrow}U(\theta,\phi)^\dagger$ and the slowest decaying mode $L_2$ of the Liouville superoperator $\mathcal{L}$ generating the dynamics,
\begin{equation}
    \label{eq:objective_function}
    \chi_2(\theta,\phi) =  \left| \Tr\left[L_2 \rho_0' \right]\right|,
\end{equation}
obtained by exact numerical diagonalization\footnote{This is not always possible, since $\mathcal{L}$ is generally not Hermitian and might require Jordan normal form.} (see~\methods{} for details on the optimization task and the sensitivity of the dynamics on $\theta$ and $\phi$.). Note that, ideally, the preprocessing sequence should be as short as possible to best approximate a local unitary. In practice, however, preprocessing times are limited by the bandwidth of amplitude and detuning pulses via a class of time-energy uncertainty relations known as quantum speed limits~\cite{Campaioli2020}. 

Following this approach, we define the fast and slow preprocessing sequences, as well as a ``control group'' sequence obtained by foregoing the preprocessing step, i.e., by setting $U(0,0) = \mathbbm{1}$. The sequences, reported in the \methods{} section, are run on both devices, measuring the atoms on the $z$-basis at some keyframes $t_k$, sampling 200 repetitions (shots) at each keyframe (See \data{} for access). 
\begin{figure*}
    \centering
    \includegraphics[width=0.90\textwidth]{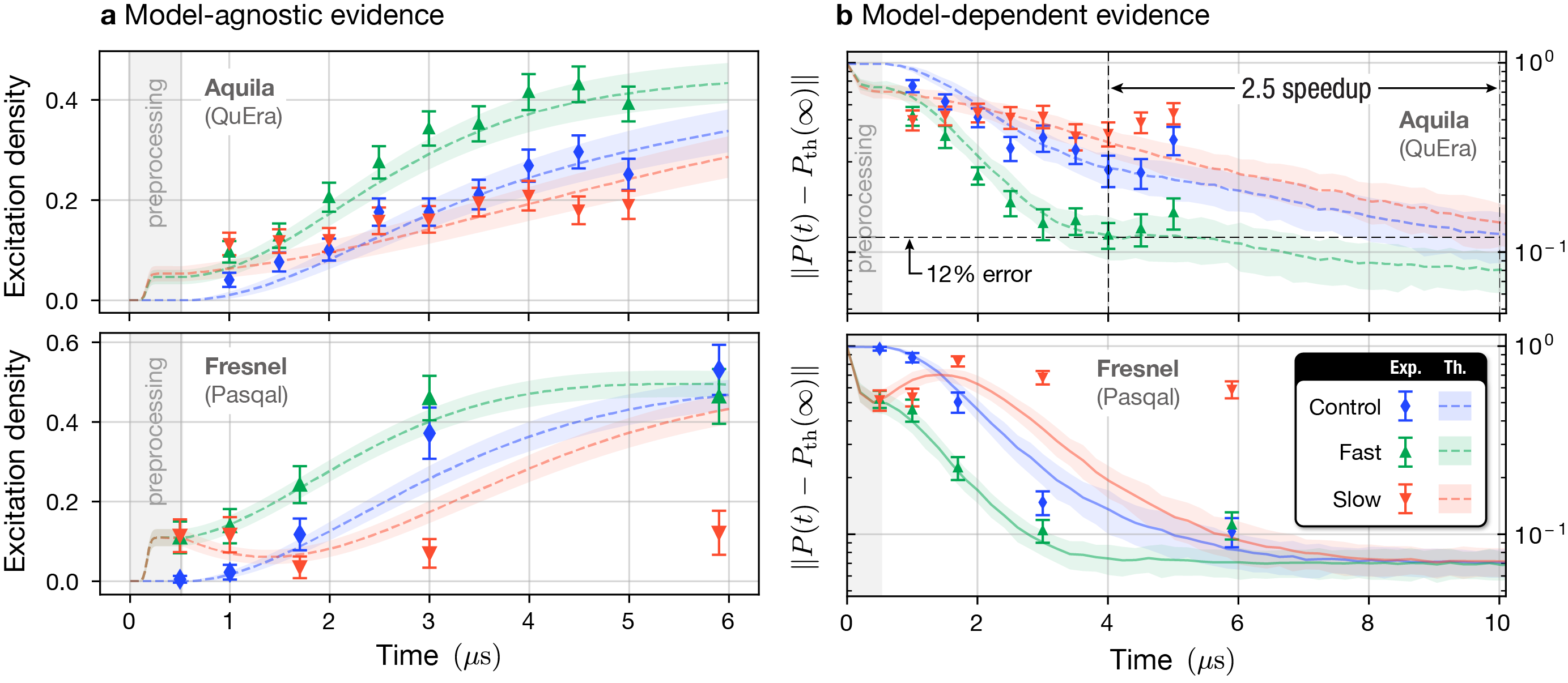}
    \caption{\textbf{Evidence of many-body quantum Mpemba effect.} Model-agnostic evidence of the QME shown in panel \figpanel{a}, by the dynamics of the excitation density $n(t)$, i.e., the average number of excitations per atom on QuEra's Aquila (top) and Pasqal's machine Fresnel (botton). Model-dependent evidence of the QME is show in panel \figpanel{b}, the dynamics of the distance from the equilibrium configuration $\|{P}(t) - {P}_\mathrm{th}(\infty)\|$, where ${P}(t)$ is the measured configuration vector at time $t$ and ${P}_\mathrm{th}(\infty)$ is the equilibrium configuration vector obtained from theory. Data points are obtained from 200 shots on each device; in~\figpanel{a} error bars represent the standard deviation, while in~\figpanel{b} they are obtained via bootstrapping, drawing 200 data replacements for 1000 times. The dashed lines represent expected excitation density dynamics obtained by exact numerical solution of the simulated sequence calibrated to each device (see Tab.~\ref{tab:device_params} and~\ref{tab:simulation_params}), with the shaded area being the standard deviation. On both devices, the fast sequence approaches equilibrium at the fastest rate, while the slow sequence is overtaken by the control sequence, despite starting from an initial higher excitation density---a signature of the Mpemba effect. In~\figpanel{b} we extrapolate the theory beyond the accessible timescale of 6~\us{ } to show the fast sequence has a 2.5-fold speedup against the control sequence when sampling equilibrium configuration within 12\% error on Aquila. We attribute deviations from theory in of the slow sequence on Fresnel to frequency drift in the control laser fields likely due to thermal fluctuations of the device~\cite{Yang2016,Proctor2020,Evered2023,Chen2025}.}
    \label{fig:mpemba_evidence}
\end{figure*}

\subsection*{Evidence of quantum Mpemba effect}
\noindent
Before discussing benchmarking and calibration results, we present evidence of the quantum Mpemba effect. We begin by analyzing the dynamics of the excitation density $n(t) := \sum_{i=1}^N \Tr[\rho(t) n_i ]/N$ on the atom ring. The results, shown in Fig.~\ref{fig:mpemba_evidence}~\figpanel{a}, offer a model-agnostic validation of speedup and slow-down with respect to the control sequence. Importantly, the slow sequence is constructed to start from a higher initial excitation density than the fast and control states. We do this to show that the relaxation rate does not depend on the initial excitation density $n(0)$, but rather on full state of the system. Note that, while the exact value of the equilibrium state's excitation density $n(\infty)$ is not experimentally accessible due to device limitations on the length of the sequences, we use best-fit model to obtain theoretical predictions for $n(\infty)\approx 0.5$ on each device. 

We then use Eq.~\eqref{eq:lindblad} to calculate the equilibrium states predicted by the theory, and evaluate the rate of approach of to equilibrium of each sequence. To do so we calculate the distance between the sampled system configurations $P(t)$ and the equilibrium configuration predicted by the theory $P_\mathrm{th}(\infty)$ (see Eq.~\eqref{eq:system_configuration} in \methods{} for details on how the system configurations are defined, measured, and calculated). The results, shown in Fig.~\ref{fig:mpemba_evidence}~\figpanel{b}, offer model-dependent evidence of the quantum Mpemba effect. 

Firstly, we note that the fast sequence approaches equilibrium at a significantly faster rate than the slow sequence. Both the fast and control sequences overtake the slow sequence within the first 2.5 \us\ on both devices, in spite of the slow sequences starting from a larger initial excitation density and smaller distance from equilibrium; this is the paradigmatic signature of the Mpemba effect. 

We can also interpret our results through the lens of equilibrium sampling, which is at the heart of applications like complex optimization~\cite{Huber1996, Lotshaw2023,Al-Kayed2025}, dissipation-driven error correction~\cite{Reiter2017}, simulation of exotic phases of matter~\cite{Islam2011,Zhang2017c,Andersen2025,teza2025finitetemperature}, and quantum Boltzmann machines~\cite{PhysRevX.8.021050}. In this context, any acceleration toward equilibrium is beneficial, reducing sequence costs and enabling tasks that would otherwise be out of reach. This speedup is particularly evident on Aquila, where the fast sequence reaches the 12\% error threshold within 4~\us. This threshold would not be accessible within the maximum schedule length available on the devices without the acceleration provided by the fast preprocessing sequence, neither for the control nor the slow sequence. We use Eq.~\eqref{eq:lindblad} to estimate that the control sequence on Aquila would reach the 12\% error threshold at time $t=10$ \us, resulting in a 2.5-fold speedup for the fast sequence against the typical relaxation rate of the system, as shown in Fig~\ref{fig:mpemba_evidence}~\figpanel{b} (top). 

Let us also note that both devices are in great agreement with theory predictions during the first 2~\us, with Aquila maintaining a good agreement across the full length of the sequences. However, the slowdown obtained on Fresnel is stronger than expected, with the data points for the slow sequence being incompatible with the model for times larger than 2 \us. By analyzing the time-dependent excitation of each individual atom (see Fig.~\ref{fig:individual_atoms} in~\methods), we can rule out atom loss from the trap as the origin of the observed discrepancy. Instead, amplitude and detuning calibration results shown in Fig.~\ref{fig:benchamrking_calibration} indicate that the effect is compatible with frequency drifts of the control lasers, most likely induced by slow thermal fluctuations, a known challenge in Rydberg-atom experiments~\cite{Yang2016,Proctor2020,Evered2023,Chen2025}. 

\subsection*{Benchmarking Rydberg-atom devices}
\noindent
We now use the local excitation measurements collected for the three sequences to benchmark the quality of each device. Several figures of merit defined on the computational basis are available for this purpose, from established ones, like the euclidean distance between classical system configurations, to recently introduced benchmarks like the \textit{quantum many-body score}~\cite{erbin2026many}.  
Here, we evaluate the \textit{Bhattacharyya coefficient} between the measured and predicted system configuration, $P$ and $P_\mathrm{th}$, respectively, 
\begin{equation}
    \label{eq:BC_coefficient}
    BC(P,P_\mathrm{th}):=\sqrt{P}\cdot \sqrt{P_\mathrm{th}},
\end{equation}
to leverage all the information gained from local measurements, where $\cdot$ denotes the inner product between the two vector and where $\sqrt{P}$ denotes the element-wise square root. Importantly, $BC$ can be understood as the classical counterpart of the quantum fidelity in the local excitation basis. As such, it defines an upper-bound on the quantum fidelity~\cite{Fuchs1996}
\begin{equation}
    \label{eq:fidelity_to_BC}
    \mathcal{F}(\rho,\rho_\mathrm{th}):=\left(\Tr\sqrt{\sqrt{\rho}\rho_\mathrm{th}\sqrt{\rho}}\right)^2\leq BC^2(P,P_\mathrm{th}),
\end{equation}
where $\rho$ and $
\rho_\mathrm{th}$ represent the full state operator of the qubit register on the device and predicted by theory, respectively. 

The results, presented in Fig.~\ref{fig:benchamrking_calibration}, show that both devices achieve good performance within the first 2~\us{}, when they operate in a largely coherent regime. Notably, both perform exceptionally well on the \textit{control} task, reaching $BC^2 \geq 0.9$ over this time window, while their performance on the \textit{fast} and \textit{slow} tasks is lower. Beyond this time window, both devices suffer a progressive loss of quality as decoherence sets in. The slow task is especially punishing, leading to a pronounced degradation on both devices, and particularly on Fresnel. This underlines the demanding nature of the slow preprocessing task, which is acutely sensitive to initial-state preparation and to frequency drifts over the course of the protocol, despite its slower relaxation towards equilibrium. These results establish that coherent-regime fidelity alone is insufficient to characterize device performance: the differentiation between devices emerges only under the combined stress of decoherence, sensitivity to state preparation, and control drifts, with the slow task acting as the decisive discriminator in our demonstration.

\subsection*{Sequence calibration}
To conclude, let us look at our results from the angle of device calibration. Most quantum computing architectures, from superconducting qubits to trapped-ions, are sensitive to variations of environmental conditions, that cause their effective control parameters to drift over the course of a sequence. In Rydberg atom platforms, temperature fluctuations are known to cause slow drifts in the frequency of the driving laser fields~\cite{Yang2016,Proctor2020,Evered2023,Chen2025}. Characterizing their magnitude and nature is a challenge that is crucial both for correcting measurement outcomes and for identifying systematic limitations that guide hardware improvements. To tackle this challenge, we propose to exploit the high-sensitivity of the quantum Mpemba effect to obtain calibration primitive based on the minimization of the loss function $\varepsilon:=1-\mathrm{FoM}$. This calibration principle can be implemented in a variety of ways, from static time-averaged corrections on the control pulses to time-resolved calibration based on optimal control routines~\cite{ROSSIGNOLO2023108782, Ansel_2024}. 
\begin{figure*}
    \centering
    \includegraphics[width=\textwidth]{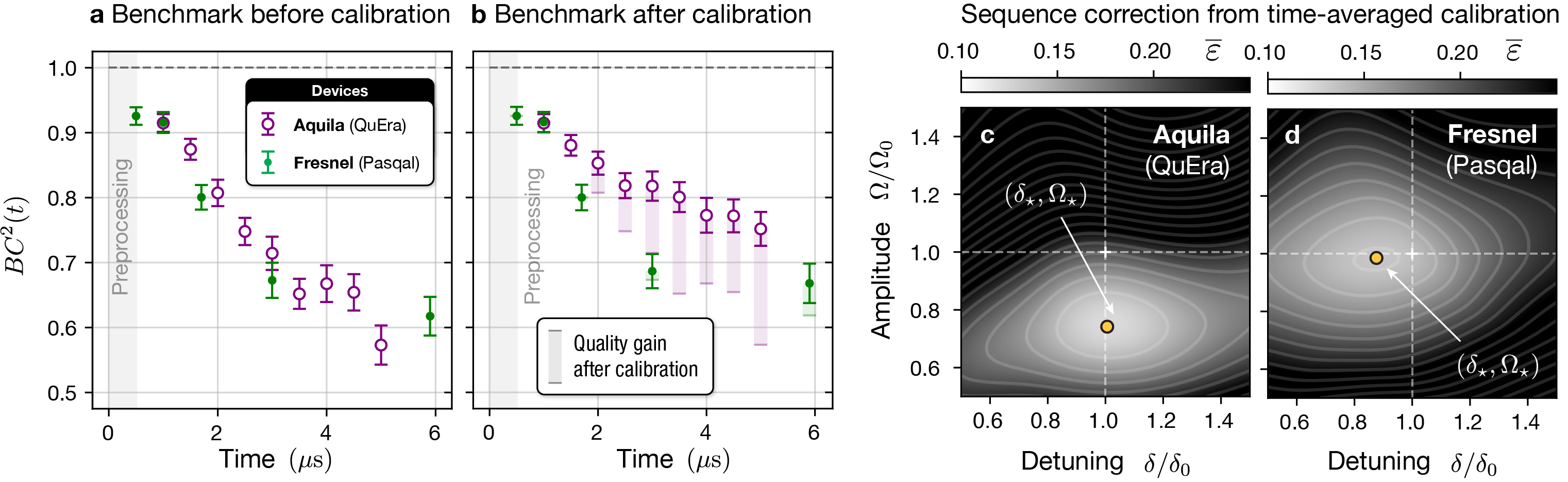}
    \caption{\textbf{Benchmark and sequence corrections from calibration.} \figpanel{a}, Benchmark score of the Aquila (QuEra) and Fresnel (Pasqal) at each keyframe averaged over control, fast, and slow sequences. Error bars denote one standard deviation from bootstrapping. \figpanel{b}, Benchmark scores after time-resolved calibration. Vertical shaded regions indicate the quality gain achieved through calibration. The modest improvement obtained on Fresnel highlights the stringency of the benchmark test, with the slow preprocessing sequence proving the most challenging to reproduce.
    Heatmap of the time-averaged error landscapes $\overline{\varepsilon}(\delta, \Omega)$ defined in Eq.~\eqref{eq:average_error} on  Aquila (\figpanel{c}) and Fresnel (\figpanel{d}), shown as a function of the deviations from the target detuning and amplitude parameters $(\delta_0,\Omega_0)$ encoded in the sequences. Our calibration routine allows us to to pinpoint the parameters $(\delta_\star, \Omega_\star)$ that minimize the error on each device,  reported in Tab.~\ref{tab:simulation_params}. See Fig.~\ref{fig:full_static_calibration} for the error landscapes of each individual sequence.}
    \label{fig:benchamrking_calibration}
\end{figure*}

We begin by performing time-averaged calibration, which model frequencies drifts as static deviation of amplitude ($\Omega$) and detuning ($\delta$) control fields from their target values $\delta$ and $\Omega$. To do so, we minimize the time-averaged loss function
\begin{equation}
    \label{eq:average_error}
    \overline{\varepsilon}(\delta, \Omega): = 1 - \frac{1}{m}\sum_{t_k} BC^2\left[P(t_k), P_\mathrm{th}(t_k; \delta, \Omega)\right],
\end{equation}
in the $(\delta, \Omega)$ plane around the target model parameters $(\delta_0, \Omega_0)$. The results, presented in Fig.~\ref{fig:benchamrking_calibration}, show the average control drifts on each device. We then perform a time-resolved calibration based on a greedy optimization; see Eq.~\eqref{eq:greedy} \methods{} for details. Applying this routine yields time-resolved calibrated pulses, which we use to re-evaluate the quality of each device at each keyframe, shown in Fig.~\ref{fig:benchamrking_calibration} and Fig.~\ref{fig:time-dependent_benchamrking_calibration} in ~\methods{}. Notably, calibration results in significant quality improvement on each device. We also notice that time-averaged and time-resolved approaches produce qualitatively consistent corrections in the control fields, as shown in Fig.~\ref{fig:calibrated_pulses} in \methods{}, confirming that the dominant drifts are slow on the timescale of the protocol. 

The key outcome of our calibration results is that Aquila and Fresnel exhibit ``complementary'' drift profiles, as shown in Fig.~\ref{fig:benchamrking_calibration} and Fig.~\ref{fig:full_static_calibration} in \methods{}. On Aquila, the detuning $\delta$ is faithfully reproduced while the amplitude $\Omega$ requires requires a 20\% re-calibration, with $\Omega_\star/\Omega_0 \approx 0.8$. On Fresnel, the situation is reversed, with $\Omega$ being well reproduced and $\delta$ requiring a $10\%$ correction, with $\delta_\star/ \delta_0\approx 0.9$. Despite sharing an almost identical architecture, the two devices are therefore limited by drifts in orthogonal control channels . The time-averaged loss landscapes of Fig.~\ref{fig:full_static_calibration} make this structure explicit. Both devices display qualitatively similar error morphologies---a general feature that stems from the underlying physics of neutral-atom arrays rather than from machine-specific conditions---yet their minima are displaced along different axes, reflecting distinct calibration errors. Furthermore, the contour lines of equal error of the fast and slow sequences intersect transversely in the ($\delta, \Omega$) plane, so that their superposition forms a sharp valley whose minimum pinpoints the calibration parameters unambiguously. 

Crucially, the ability to disentangle errors in these two control channels is a direct consequence of probing complementary relaxation regimes, i.e., the fast and slow sequences, which respond differently to perturbations in $\delta$ and $\Omega$, breaking degeneracies that either sequence alone would leave unresolved.

Several broader implications follow. First, anomalous relaxation broadens the useful window for calibration by simultaneously probing the fastest and slowest relaxation rates accessible to the device, a range that could be extended further through optimal-control state preparation. Second, although the absolute calibration corrections are device- and session-dependent, the morphology of the loss landscapes appears to be generic, reflecting the underlying physics of neutral-atom arrays rather than idiosyncrasies of a particular device or execution. Third, the collective character of many-body relaxation offers a scalable route to amplify calibration sensitivity: by engineering spectral gaps $\mathrm{Re}[\lambda_3]-\mathrm{Re}[\lambda_2]$ that grow with system size $N$, Mpemba-based diagnostics can extend beyond the single-qubit and two-qubit regime in which many current calibration protocols operate.

\subsection*{Conclusions}
\noindent
In this work we introduced a benchmarking and calibration primitive for quantum computing and simulation. The approach, based on the quantum Mpemba effect, leverages high-sensitivity to initial conditions and noise to define a stringent quality test and a precise calibration routine. We demonstrated the use of this approach by implementing accelerated and slowed-down relaxation in a six-atom ring on two leading Rydberg-atom quantum simulators, QuEra's Aquila and Pasqal's Frensel, showing that relaxation rates depend on the full state of the system, rather then just its projection onto the observable ground-Rydberg basis. This is one of the signatures of the quantum Mpemba effect, and its first demonstration in many-body open quantum system, here shown in an unprecedented simulation across multiple devices. We then used the measurement outcomes to benchmark and calibrate the devices, identifying different systematic errors in the control fields of Aquila and Fresnel, demonstrating the prescriptive power of the many-body quantum Mpemba effect.

Our results position Mpemba dynamics as a unifying framework connecting non-equilibrium many-body physics with the practical validation and calibration of quantum hardware.
By directly linking relaxation behavior to device performance, this approach provides a route to assess and compare quantum simulators in regimes that are both physically relevant and classically intractable.
At a time when quantum platforms are rapidly advancing yet remain difficult to benchmark reliably, such dynamical and application-oriented validation tools are becoming essential.
In this context, the ability to extract meaningful performance metrics and calibration information from simple, scalable protocols opens new possibilities for systematic cross-device comparison and optimization.
More broadly, our work points towards the development of standardized validation procedures for quantum simulators, enabling robust and reproducible exploration of many-body quantum phenomena and strengthening the foundations for future practical applications.

Looking forward, a central goal is to integrate this approach into validation of quantum devices across all platforms at scale. For Rydberg atom quantum simulators, our method is scalable by design: the transverse-field Ising model with pure local dephasing in the computational basis admits an exact solution for the spectrum and eigenmodes of relaxation~\cite{Zheng2023exact, Prosen_2008, Roberts2023, Foss-Feig2013}, and weak local relaxation can be incorporated perturbatively. This allows us to evaluate the quality of neutral atom devices on both fast and slow dynamics in arbitrarily large registers, using only frugal local measurements. The open challenge is to extend the routine across the broader landscape of quantum devices---superconducting qubits, trapped ions, and photonic platforms---each with its own decoherence channels and its own relaxation physics to exploit. We expect the quantum Mpemba effect to have a central role in tackling this challenge, by providing a universal diagnostic tool for the development of quantum technologies in the coming decade.

\begin{acknowledgments}
\noindent
The authors thank Simone Montangero, the Quantum Computuing and Simulation Centre (QCSC), Daniele Ottaviani, and CINECA, for the computing resources on Pasqal's Fresnel Machine, as part of the ISCRA-C EmpeRyd Project (HP10CML0M6). This work was supported by resources provided by the Pawsey Supercomputing Research Centre. It was carried out within the Pawsey Supercomputing Research Centre’s Quantum Supercomputing Innovation Hub, made possible by a grant from the Australian Government through the National Collaborative Research Infrastructure Strategy (NCRIS). Computational resources were provided by the Pawsey Supercomputing Research Centre’s Setonix Supercomputer~\cite{pawsey_setonix_2023}. F.C. thanks Constantin Dalyac, Charles Moussa, Roland Guichard and Pasqal for technical support. The authors also thank Tommaso Macrì, Jonathan Wurz, QuEra, Pascal Elhai, and Pawsey Centre for support and quantum computing resources on QuEra's machine Aquila. F.C. thanks Simone Notarnicola and A. G. Catalano for discussions on neutral atom quantum devices.  F. C. acknowledges that results incorporated in this standard have received funding from the European Union Horizon Europe research and innovation program under the Marie Skłodowska-Curie Action for the project SpinSC. O.R. acknowledges financial support from ISF Grant No. 232/23 and from the Minerva foundation with funding from the Federal German Ministry for Education and Research. 
\end{acknowledgments}

\subsection*{Data and Code availability}
\hypertarget{data}{}
\noindent
The data and the code supporting the findings in this work are available on Zenodo at \href{https://doi.org/10.5281/zenodo.20423683}{10.5281/zenodo.20423683}.

\subsection*{Contributions}
G.T. conceived the idea of using anomalous many-body relaxation phenomena in quantum simulators as validation and calibration tools for benchmarking quantum devices and led the project's development. F.C. developed the framework for benchmarking and calibration, implemented its experimental realization on the quantum hardware, including the mapping of the model onto the devices, the optimization and execution of the pulse sequences, and the numerical simulations.
All authors discussed and analyzed the results and contributed to writing the manuscript.

\section*{Methods}
\hypertarget{methods}{}
\renewcommand{\theequation}{S.\arabic{equation}}

\subsection*{Anomalous relaxation}
Consider an open quantum systems of dimension $d$, whose dynamics is governed by a Markovian quantum master equation $\dot{\rho}_t:= \frac{d\rho_t}{dt} = \mathcal{L}[\rho]_t$ in the Gorini-Kossakowski-Sudarshan-Lindblad (GKSL) form of Eq.~\eqref{eq:lindblad}, where $\rho_t$ is the density operator representing the state of the system at time $t$, $\mathcal{L}$ is the time-independent Liouville superoperator generating the dynamics, $H$ is the Hamiltonian of the system, responsible for coherent evolution, and $J_j$ are quantum jump operators, modeling incoherent transitions~\cite{Breuer2002}, with $\hbar\equiv 1$. Eq.~\eqref{eq:lindblad} provides the most general form for a completely positive and trace-preserving Markovian quantum master equation, being sufficiently general for applications in a vast range of fields, from neutral atoms to electron spin resonance~\cite{Campaioli2024a}.
\begin{figure}
    \centering
    \includegraphics[width=0.48\textwidth]{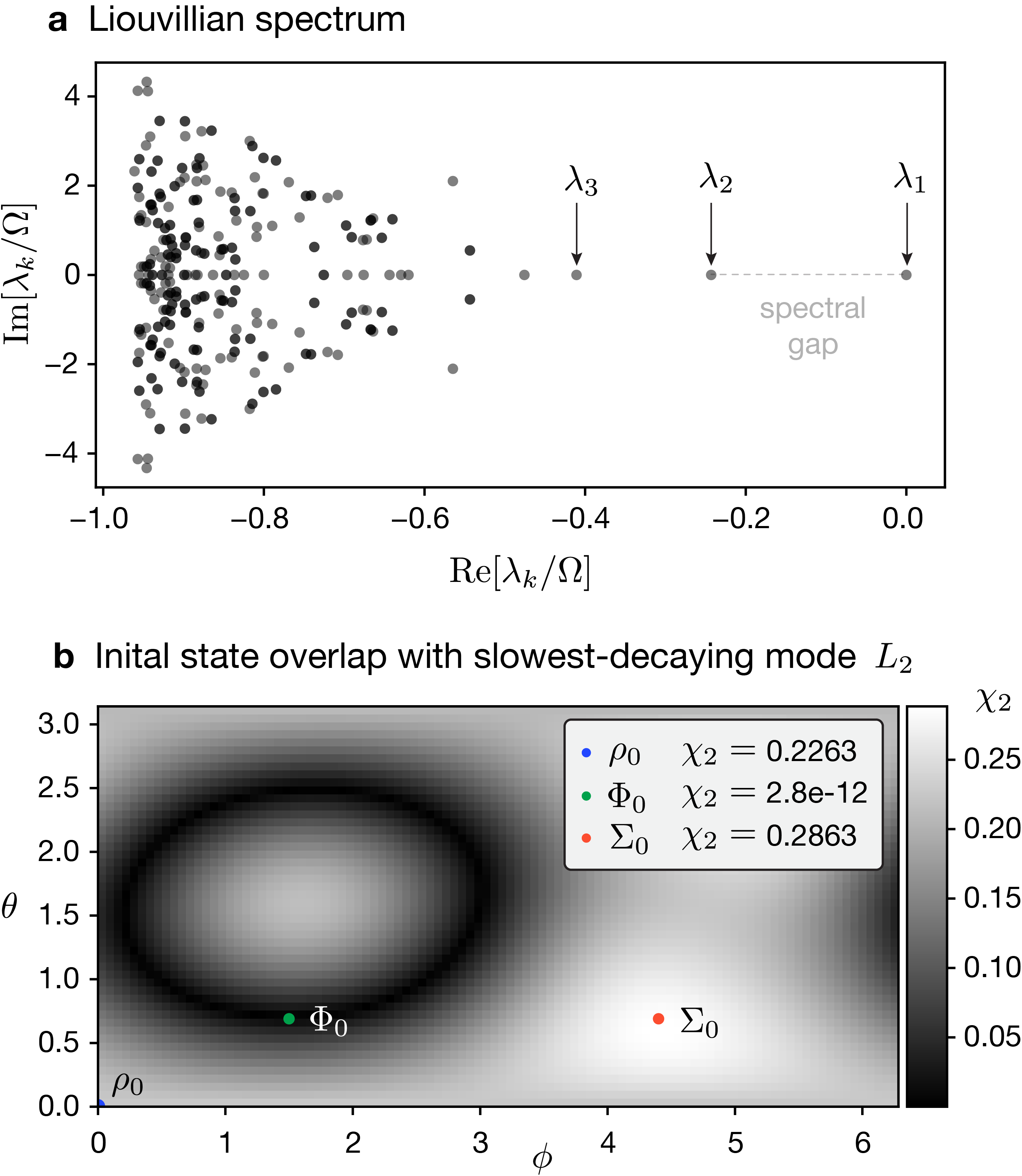}
    \caption{\textbf{Tuning relaxation by controlling the overlap with the slowest-decaying mode.} \figpanel{a}, The first 350 elements of the spectrum $\{\lambda_k\}_{k=1}^{d}$ of the Liouville superoperator $\mathcal{L}$ associated with the Hamiltonian of Eq.~\eqref{eq:SimulatorHamiltonian}, with $N=6$, $\Omega = 0.584$, $\delta = 0.866$, $C_6/r_0^6 = 0.866$, $\gamma_0=0.222$, $\gamma_\downarrow=0.01$, plotted in the complex plane relative to the system's transverse field amplitude $\Omega$. \figpanel{b}, Overlap $\chi_2(\theta, \phi)$ between the preprocessed initial state $\rho_0'(\theta,\phi)$ and the slowest decaying mode $L_2$ for $N=6$. The fiducial initial state $\rho_0 = \ketbra{\downarrow}{\downarrow}^{\otimes N}$ and the slow-decaying state $\Sigma_0$ have a significant overlap $\chi_2 > 0.2$, while the fast-decaying state $\Phi_0$ has an almost vanishing overlap $\chi_2  = 2.81 \cdot 10^{-12}$. Note that $\Phi_0$ and $\Sigma_0$ are approximated on the simulator, due to the finite bandwidth of the amplitude and detuning pulses. This approximation can, in principle, be improved using quantum optimal control methods~\cite{Koch2022}.}
    \label{fig:spectrum_and_optimisation}
\end{figure}

The superoperator $\mathcal{L}$ is generally not Hermitian and, therefore, it may not be diagonalizable and require Jordan normal form.
When the superoperator $\mathcal{L}$ can be diagonalized, it can be represented by its complex spectrum $\{\lambda_k\}_{k=1}^{d^2}$, which we assumed to be ordered according to their real part so that $\Re[\lambda_k]\geq\Re[\lambda_{k+1}]$, and the associated left and right eigen-matrices $L_k$ and $R_k$, respectively, here normalized via $\Tr[L_k R_{k'}]=\delta_{kk'}$. Note that this normalization still leaves freedom: 
if one multiplies $L_k$ by an arbitrary nonzero scalar $\alpha_k$ and $R_k$ by $\alpha_k^{-1}$ the normalization condition does not change. 
Note that $\lambda_1=0$ is associated with the steady state $\rho_\mathrm{ss} = R_1$ via the additional normalization $\Tr[\rho_\mathrm{ss}]=1$, such that $\mathcal{L}[R_1]=0$. 
Through this decomposition, Eq.~\eqref{eq:lindblad} can be integrated to the solution
\begin{equation}
    \label{eq:solution}
    \rho_t = \rho_\mathrm{ss} + \sum_{k=1}^{d^2} c_k e^{t\lambda_k} R_k,
\end{equation}
where $c_k:=\Tr[\rho_0 L_k]$ are amplitudes (overlaps) between the initial state $\rho_0$ and the relaxation modes $L_k$.
If the steady state $\rho_\mathrm{ss}$ is unique, $\Re[\lambda_2]<0$ determines the so-called \textit{spectral gap} of the superoperator, setting the slowest relaxation rate of the dynamics. Accordingly, for $t\to\infty$, the system approaches equilibrium as
\begin{equation}
    \label{eq:approach_to_ss}
    \lim_{t\to\infty}\lVert \rho_t-\rho_\mathrm{ss}\rVert \sim \exp\left(\Re[\lambda_2]t\right),
\end{equation}
as long as $c_2\neq 0$, with $\lVert\cdot \rVert$ being the Frobenius norm~\cite{Macieszczak2016} 

To control relaxation rates via initial state preparation, or \textit{preprocessing}, one can tune the overlap between the initial state of the system and the slow-decaying modes of the Liouville superoperator. For instance, to accelerate relaxation one can prepare some initial state $\rho_0$ with no overlap with the slowest decaying mode\footnote{It is important to note that Eq.~\eqref{eq:c2=0} sets a condition that is relevant for ``long-time behavior'' of the system dynamics, and might not be the right choice if one wants to achieve acceleration within a specific transient timescale.} $L_2$
\begin{equation}
    \label{eq:c2=0}
    c_2 = \Tr[\rho_0 L_2]=0.
\end{equation}
The ability to satisfy Eq.~\eqref{eq:c2=0} exactly depends on the limitations of the control operations available for the preprocessing step~\cite{Carollo2021,Kochsiek2022}. Let us assume that we can prepare a slow decaying state $\Sigma_0$ such that $\Tr[\Sigma_0 L_2]\neq0$ and a fast decaying mode $\Phi_0$ such that $\Tr[\Phi_0 L_2]= 0$ and 
\begin{equation}
    \label{eq:initial_distances}
    \lVert \Sigma_0 - \rho_\mathrm{ss} \rVert < \lVert \Phi_0 - \rho_\mathrm{ss} \rVert,
\end{equation}
Eq.~\eqref{eq:initial_distances} states that the fast-decaying state $\Phi_0$ is further away from the steady state than the slow-decaying state $\Sigma_0$ for some notion of distance. This means that after a sufficiently long time $\tau$, the fast-decaying state $\Phi_t$ overtakes $\Sigma_t$ in its approach to equilibrium, $\lVert \Phi_\tau - \rho_\mathrm{ss} \rVert < \lVert \Sigma_\tau - \rho_\mathrm{ss} \rVert$ and eventually approach the steady state as $\lim_{t\to\infty}\lVert \Sigma_t - \rho_\mathrm{ss}\rVert \sim \exp\left(\Re[\lambda_3]t\right)$, assuming that $\Tr[\Phi_0 L_3] \neq 0$. This ``counterintuitive'' crossover in the approach to equilibrium is one of the signatures of quantum Mpemba effects~\cite{Nava2024, Teza2026, Ares2025}, and will be the focus of our demonstration on a neutral atom programmable quantum simulator.

\subsection*{Model definition}
To demonstrate control over relaxation rates on Aquila and Fresnel, we implement a variant of the transverse field quantum Ising Hamiltonian of Ref.~\cite{Kochsiek2022}. The model that we implement is entirely defined by the target Hamiltonian $H$
\begin{equation}  
    \label{eq:target_Hamiltonian}
    H = \Gamma \sum_{i=1}^N \sigma_i^{(x)} + \sum_{i<j}^N V_{ij}^\pdagger \sigma_i^{(z)}\sigma_i^{(z)}, 
\end{equation}
and the local jump operators $\{J_i^{0}, J_i^{\downarrow} \}$ defined in Eq.~\eqref{eq:relax} and~\eqref{eq:deph} in the main text,
where $\sigma_i^{(\nu)}$, $\nu=x,y,z$, are Pauli operators acting on the $i$-th atom, $\Gamma$ is the transverse field strength, $V_{ij}^\pdagger$ is the distance-dependent interaction strength between atoms $i$ and $j$, $\gamma_\downarrow$ is the local relaxation rate, and $\gamma_0$ is the local dephasing rate on the ground-Rydberg basis.

\subsection*{Mapping to device Hamiltonian}
\begin{figure*}
    \centering
    \includegraphics[width=0.70\linewidth]{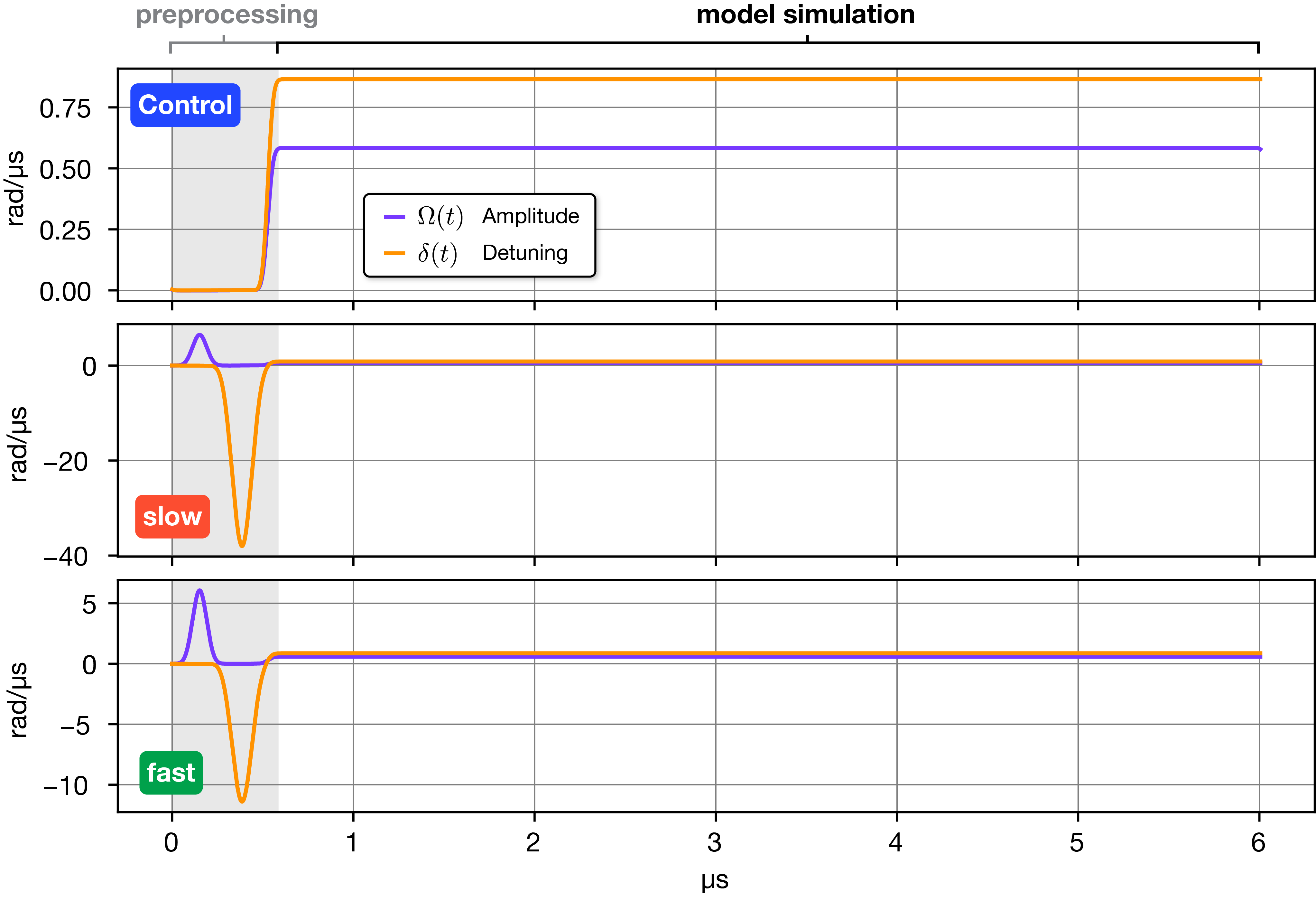}
    \caption{\textbf{Sequences, preprocessing, and model simulation.}  The time-dependent sequences for amplitude, $\Omega(t)$, and detuning, $\delta(t)$, used for the preprocessing and model simulation of the control, slow, and fast sequences, respectively, as indicated in the legend. After the preprocessing step, all sequences are set to simulate the same model, by setting amplitude and detuning to the same value and keeping them constant over time.}
    \label{fig:sequences}
\end{figure*}

The model defined by Eqs.~\eqref{eq:target_Hamiltonian},~\eqref{eq:relax} and~\eqref{eq:deph} can be accurately implemented on any globally-addressable Rydberg-atom simulator by operating the device in the ground-Rydberg ($\mathrm{gR}$) interaction~\cite{Silverio2022pulseropensource}, governed by the Hamiltonian of Eq.~\eqref{eq:SimulatorHamiltonian}. The device-specific coupling constant $C_6$ is shown in Tab.~\ref{tab:device_params} for both devices. The atoms in the devices are also affected by local relaxation and dephasing as in Eq.~\eqref{eq:relax} and~\eqref{eq:deph}, respectively. The dephasing rates are device-specific, as reported in Tab.~\ref{tab:device_params}. 

\begin{table}
\begin{tcolorbox}[tabularx={l|c|c|c|c},
boxrule=0.5pt]
\textbf{Device} & $r_0$ (\um) & $C_6$ ($\mathrm{MHz} \cdot \mathrm{\mu m}^6$) & $\gamma_\downarrow$ ($\mathrm{MHz}$) & $\gamma_0$ ($\mathrm{MHz}$)\\
\hline
Aquila & 13.58 & 5420503.0  & 0.013 & 0.10782  \\
\hline
Fresnel & 13.58 & 5410768.9 & 0.010 & 0.22222  
\end{tcolorbox}
\caption{\textbf{Device parameters.} Device-specific parameters used to implement the 6-atom register described in Fig.~\ref{fig:overview}, to obtain the same nearest-neighbor coupling $C_6/r_0^6 = 0.86(4)\pm0.001$ MHz on both devices. The relaxation and dephasing rate parameters $\gamma_\downarrow$ and $\gamma_0$, respectively, cannot be programmed. Their values are fitted to the data starting from the rates reported by QuEra and Pasqal, with an excellent agreement to their nominal value~\cite{wurtz2023aquila, Silverio2022pulseropensource}.}
\label{tab:device_params}
\end{table}

To implement Eqs.~\eqref{eq:target_Hamiltonian}---\eqref{eq:deph} under the limitations of global addressing, we prepare 6-atom ($N=6$) rings with periodic boundary conditions, placed at distances so that the next-to-nearest-neighbor coupling strength is less than 10\% of the dominant nearest-neighbor coupling strength $C_6/r_0^6$, with $r = r_{i,i+1 \mod N}$. Periodic boundary conditions are necessary to compensate for the additional $\sigma_i^{(z)}$ terms arising from $n_i n_j$ interactions in the absence of direct local addressing. Under these assumptions, the models map to each other via
\begin{equation}
    \label{eq:mapping}
    (\Omega, \delta, C_6/r_0^6, \gamma_0, \gamma_\downarrow) \to (2\Gamma, 4V, 4V, \gamma_0, \gamma_\downarrow).
\end{equation}

\subsection*{Preprocessing and sequence definition}
When performing sequences on neutral atom simulators, the dominant budget constraint is the total number of available shots. Each shot corresponds to a single preparation and measurement cycle, which needs to be repeated hundreds of times to build reliable statistics, depending on the size of the register, with larger registers requiring more shots. Therefore, a cost-effective and convincing demonstration of controlled relaxation requires careful choice of system parameters to achieve sufficiently different relaxation rates and pin the relaxation crossover of the quantum Mpemba effect. This task is further complicated by stringent constraints on the available device, restricting atoms' placement on a pre-calibrated layout on Pasqal's Fresnel, as well as minimum and maximum pulse duration and amplitude, and fixed relaxation rates. 

To select ideal simulation parameters we numerically diagonalise the superoperator $\mathcal{L}$ associated with the considered model for $N=6$ atoms in an hexagonal ring with fixed $\gamma_\downarrow$, $\gamma_0$, and interaction strength $V$, to obtain the ordered spectrum $\{\lambda_k\}_{k=1}^{d^2}$ and eigenmodes $L_k$, $R_k$. We then look for the fast-decaying (slow-decaying) state $\Phi_0$ ($\Sigma_0$) by numerically minimizing (maximizing) the function $\chi_2:=|c_2|$, where
\begin{equation}
    \chi_k(\theta,\phi) =  \left| \Tr\left[L_k U(\theta,\phi)\rho_0U(\theta,\phi)^\dagger \right]\right|,
\end{equation}
with $\theta\in(0,\pi/2], \phi\in(0,2\pi]$,
where $\rho_0:=\otimes_{i=1}^N\ketbra{\downarrow}{\downarrow}_i$ is the fiducial initial pure state of the register, for which every atom is in the ground state $\ket{0}_i$ of the local ``longitudinal'' field $\sigma_i^{(z)}$, where
\begin{equation}
    \label{eq:local_unitary}
    U(\theta,\phi)=\bigotimes_{i=1}^N Z_i(\phi)X_i(\theta),
\end{equation}
and where $Z_i(\phi)$ and $X_i(\theta)$ are given in Eq.~\eqref{eq:Z} and~\eqref{eq:X}, respectively. 
Locally, the preprocessed initial state takes the form
\begin{equation}
    \ket{\theta,\phi}_i:=i\sin\frac{\theta}{2}e^{i\phi/2}\ket{\downarrow}_i+\cos\frac{\theta}{2}e^{-i\phi/2}\ket{\uparrow}_i,
\end{equation}
which is associated with the local Bloch vector for qubit $i$ given by $v(\theta,\phi) = (-\sin\theta\sin\phi,-\sin\theta\cos\phi,-\cos\theta)$. We then determine model parameters ($\Gamma$, $V$) with suitable $\Phi_0$, $\Sigma_0$ that satisfy Eq.~\eqref{eq:initial_distances} as well as
\begin{align}
    \label{eq:vanishing_overlap}
    & |\Tr[\Phi_0 L_2]| < \epsilon \ll |\Tr[\Sigma_0 L_2]|, \;\textrm{with} \; \epsilon \approx 10^{-8},
\end{align}
while also displaying a large spectral gap
\begin{equation}
    \label{eq:spectral_gap}
    \Delta_{2,3}:=|\Re[\lambda_3]-\Re[\lambda_2]|.
\end{equation}
These conditions guarantee that the fast-decaying state has negligible overlap with the slow-decaying mode and sufficient acceleration to equilibrium.

We then simulate the system's dynamics by applying Eq.~\eqref{eq:solution} for $\rho_0$, $\Phi_0$, and $\Sigma_0$, using exact numerical methods. These have been done with both QuTiP~\cite{Johansson2012} and QuEra's Bloqade and Pasqal's Pulser open source software~\cite{Silverio2022pulseropensource}. See~\data{} for the code used for the numerical simulations. This step is essential to determine the timescale necessary to see the quantum Mpemba crossover. This is crucial due to the limited 5---6 \us{} window available for simulation on these devices. From this analysis, we obtain the simulation parameters reported in Tab.~\ref{tab:device_params}.
The sequences, shown in Fig.~\ref{fig:sequences}, are available online as reported in~\data.

\begin{table}
\begin{tcolorbox}[tabularx={l|Y|Y|Y|Y}, boxrule=0.5pt]
{} & \multicolumn{2}{c|}{Target model} & \multicolumn{2}{|c}{Calibrated model} \\
\hline
\textbf{Device} & $\Omega_0$ & $\delta_0$ & $\Omega_\star$ & $\delta_\star$\\
\hline
Aquila & 0.58(4) & 0.86(4) & 0.46(3) & 0.84(9) \\
\hline
Fresnel & 0.58(4) & 0.86(5) & 0.57(2) & 0.74(4)  
\end{tcolorbox}
\caption{\textbf{Target and calibrated model parameters.} All values are expressed in MHz. The target value are encoded in the sequences, shown in Fig.~\ref{fig:sequences}, after the preprocessing step. The calibrated values are obtained according to the time-averaged calibration procedure of Eq.~\eqref{eq:average_error} described in the main text, to minimize the error between the data and the model prediction.}
\label{tab:simulation_params}
\end{table}

\subsection*{Observables}
The modest system size $N=6$ allows us to numerically access the full information associated with its quantum state, represented by the density operator $\rho_t$. On the simulation device, instead, we can only directly access local measurements of the atoms' excitation at the end of a sequence. This corresponds to projecting the state of the system on to the full ground-Rydberg basis,
\begin{equation}
    \label{eq:local_basis}
    \mathcal{B}:=\left\{\bigotimes_{i=1}^N \ket{s_i}_i\right\},
\end{equation}
with $\ket{s_i} \in \mathcal{B}:= \{\ket{\downarrow}_i,\ket{\uparrow}_i\}$ being the ground and Rydberg state of the $i$-th atom, respectively,
consisting of all $2^N$ combinations of local products of the ground and Rydberg states. Indirectly, it is also possible to further characterize $\rho_t$, for example via local state tomography or classical shadows~\cite{qadence2024pasqal}. 

By accumulating a sufficient number of repeated measurements, or shots, we can approximate the diagonal $P(t) := \mathrm{diag}\rho(t)$ of the density operator expressed in the basis of Eq.~\eqref{eq:local_basis}, which we can use to estimate the expectation value of observables such as the excitation density. Note that the vector 
\begin{equation}
    \label{eq:system_configuration}
    P(t) = \begin{pmatrix}
        & P_1(t) \\
        & \vdots \\
        & P_{2^N}(t)
    \end{pmatrix},
\end{equation} 
has $2^N$ components (one for each element in the basis $\mathcal{B}$ of Eq.~\eqref{eq:local_basis}) and  corresponds to the system classical ``configuration'' at time $t$ with respect to the ground-Rydberg basis, which we use to estimate the distance from the theoretical equilibrium configuration,
\begin{equation}
    \label{eq:equilibrium_configuration}
    P_\mathrm{th}(\infty) = \mathrm{diag}\lim_{t\to\infty}\rho_t,
\end{equation}
which we obtain by solving for the steady state of Eq.~\eqref{eq:lindblad}.

\subsection*{Calibration}
\begin{figure*}[t]
    \centering
    \includegraphics[width=0.9\textwidth]{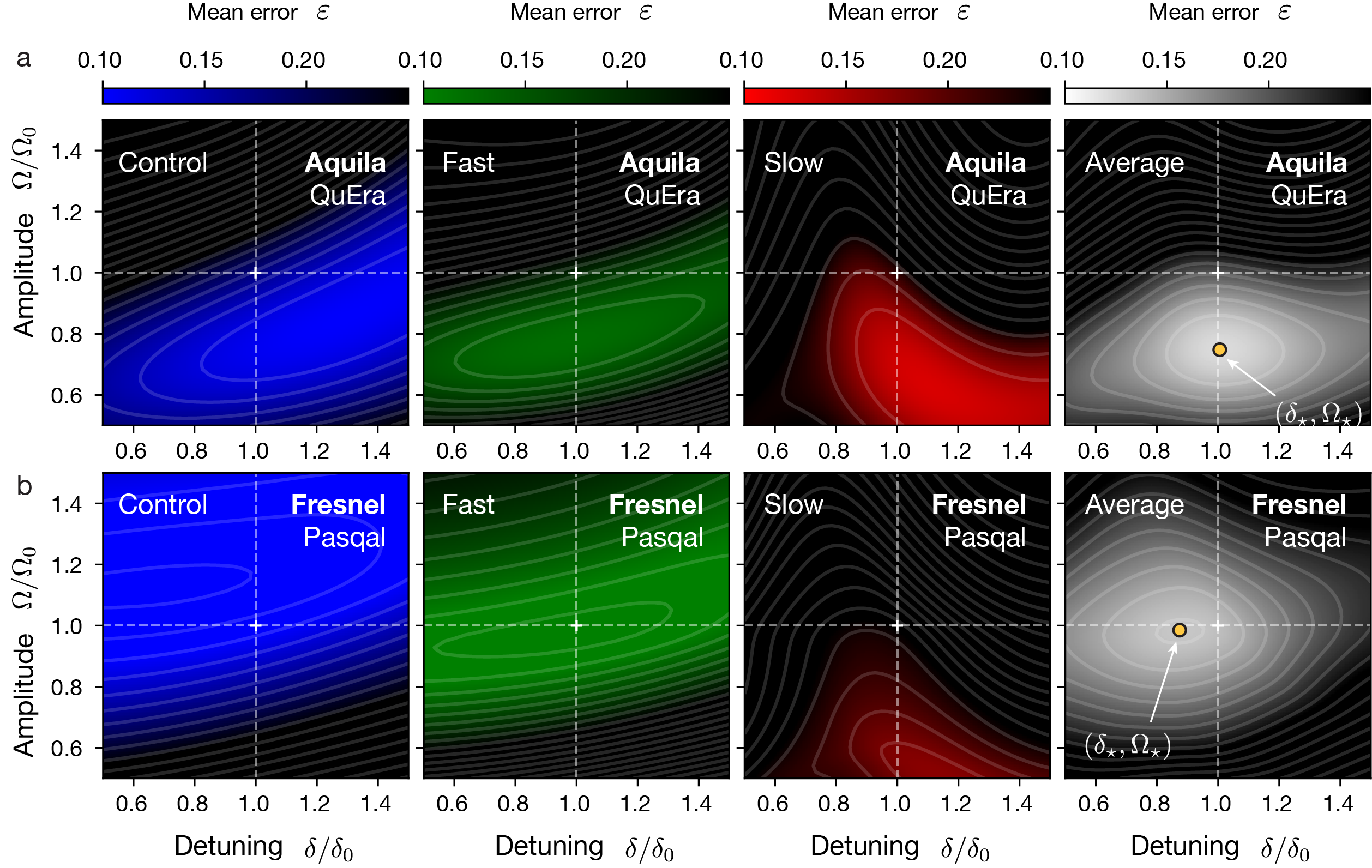}
    \caption{\textbf{Error landscape for time-averaged calibration.}  Heatmap of the mean error landscapes $\overline{\varepsilon}(\delta, \Omega)$ defined in Eq.~\eqref{eq:average_error} for each control (blue), fast (green) and slow (red) sequences, as well as for their average (white), on both Aquila (\figpanel{a}) and Fresnel (\figpanel{b}). The landscapes are shown as a function of the deviations from the target detuning and amplitude parameters $(\delta_0,\Omega_0)$ encoded in the sequences. By averaging the error landscapes across the control, fast, and slow sequences, we pinpoint the calibration parameters $(\delta_\star,\Omega_\star)$ the minimize the error on each device. This operation is facilitated by the crossing contour lines of the error landscapes of slow and fast sequences, highlighting the power of anomalous relaxation.
    }
    \label{fig:full_static_calibration}
\end{figure*}

Time-averaged calibration is performed by minimizing the loss function defined in Eq.~\eqref{eq:average_error} around the target detuning and amplitude $(\delta_0, \Omega_0)$. The results for each individual preprocessing sequence and their average are shown in Fig.~\ref{fig:full_static_calibration}.

To perform time-resolved calibration, we use a greedy minimization method. Starting from the first measurement keyframe $t_1$, we minimize the objective function
\begin{equation}
    \label{eq:greedy}
    O(t_k):=\underbrace{1-BC^2\left[ P(t_k), P_\mathrm{th}(t_k)\right]}_{\text{loss function $\varepsilon(t_k; \delta, \Omega)$}} + \underbrace{p \left[\delta^2(t_k)+\Omega^2(t_k)\right]}_{\text{penalty}},
\end{equation}
at each measurement keyframe $t_k \in \{t_1, \cdots, t_m\}$ for the family of time-dependent pulses $\delta(t)$ and $\Omega(t)$ given by piece-wise linear ramps of the form
\begin{equation}
    \label{eq:piece-wise_linear_ramps}
    f(t):= f(t_k) + \frac{f(t_{k+1})-f(t_k)}{t_{k+1}-t_k}(t-t_k), \quad t\in[t_k, t_{k+1}],
\end{equation}
with  $k = \{1, \cdots, t_m\}$. The penalty term, weighted by $p$, favors calibrated pulses that remain close to the target sequence and suppresses spurious minima far from the nominal operating point. Applying this routine to the loss averaged over the control, fast, and slow sequences, we obtain the time-resolved calibrated pulses shown in Fig.~\ref{fig:calibrated_pulses} of the \methods{}, and use them to re-evaluate the quality of each device, as shown in Fig.~\ref{fig:benchamrking_calibration} and in Fig.~\ref{fig:time-dependent_benchamrking_calibration} for each sequence. 

\begin{figure}
    \centering
    \includegraphics[width=0.48\textwidth]{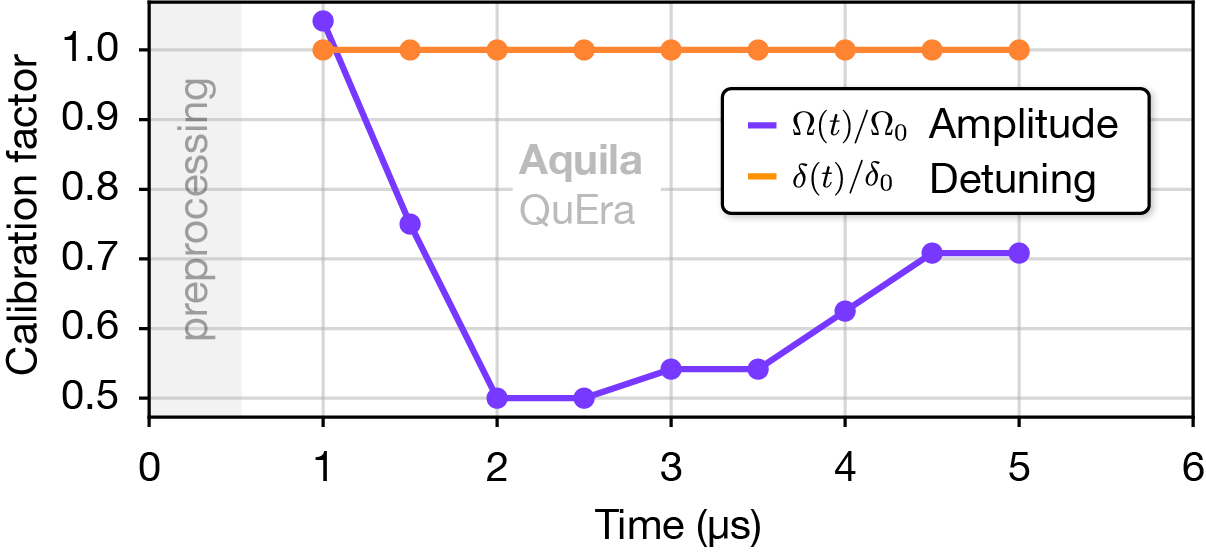}
    \caption{\textbf{Time-resolved drifts from calibration.} Slow drifts in the control parameters $\delta$ and $\Omega$, here expressed as a calibration factor from the target values $\delta_0$ and $\Omega_0$, are shown for QuEra's Aquila. The calibrated sequences are obtained using the greedy optimization for the time-resolved calibration methods defined in Eq.~\eqref{eq:greedy}.}
    \label{fig:calibrated_pulses}
\end{figure}

\begin{figure*}[ht]
    \centering
    \includegraphics[width=0.98\textwidth]{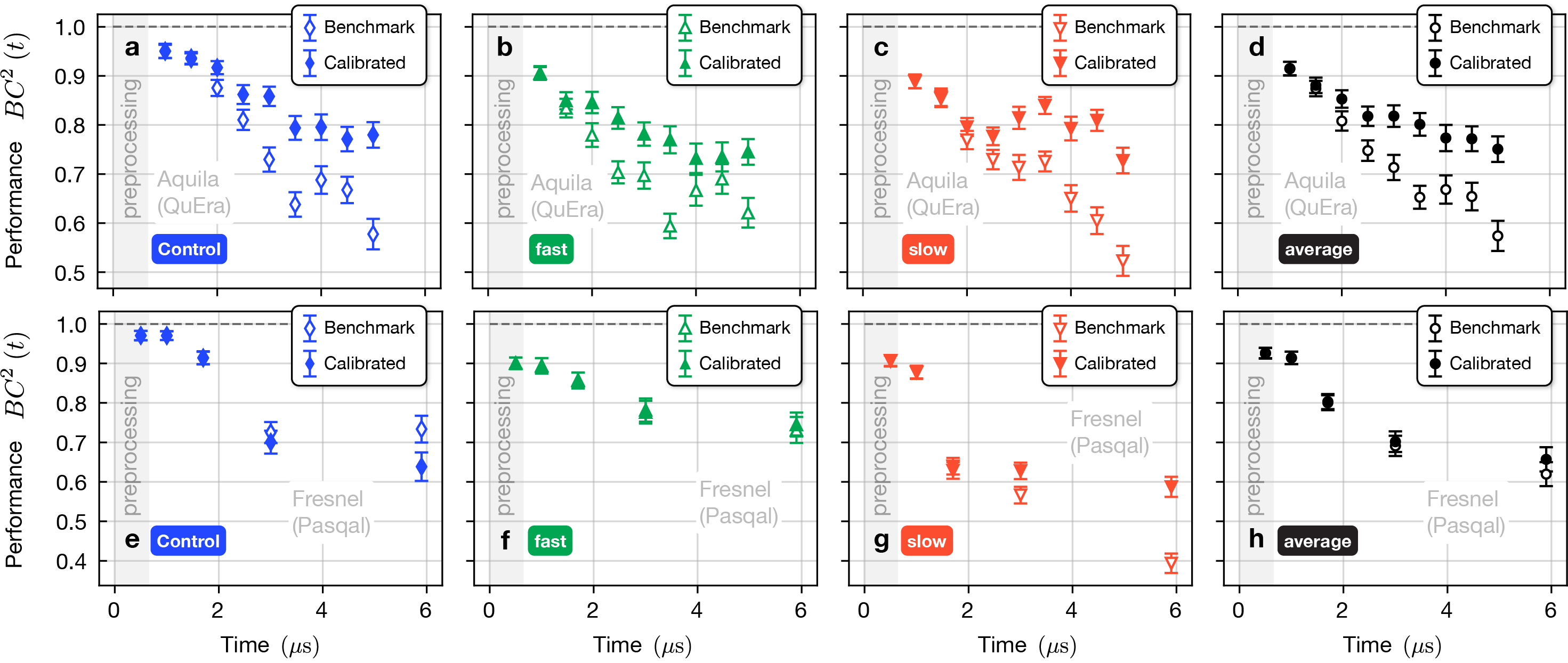}
    \caption{\textbf{Benchmarking and calibration results.} Benchmark (open markers) and calibrated (filled markers) results for Aquila (top row) and Fresnel (bottom row) under three preprocessing regimes, control (blue, \figpanel{a}, \figpanel{e}), fast (green, \figpanel{b}, \figpanel{f}), and slow (red, \figpanel{c}, \figpanel{g}), evaluated via the \textit{Bhattacharyya coefficient} ($BC$) of Eq.~\eqref{eq:BC_coefficient}. The average benchmark and calibration (black, \figpanel{d}, \figpanel{h}) is obtained as the average $BC^2$  over the three sequence. Across all conditions, calibration aims to shifts performance toward higher values relative to the uncalibrated benchmark, with error bars denoting one standard deviation from bootstrapping.}
    \label{fig:time-dependent_benchamrking_calibration}
\end{figure*}

\subsection*{Single-atom excitation dynamics}
To shed light into the discrepancy between theory and simulation results for the slow sequence on Pasqal's Fresnel machine, we look at the time-dependent excitation of each individual atom, to reveal whether any atom leaves the trap during the sequence. The results, shown in Fig.~\ref{fig:individual_atoms}, show that all the atoms have a non-vanishing excitation for all times, suggesting that no atom definitively leaves the trap at some point during the sampled shots of the sequence. These results further support our interpretation that deviations are caused by the frequency drifts shown in Fig.~\ref{fig:calibrated_pulses}. The lack of compatibility between data and theory for the slow sequence on Pasqal's Fresnel can be indeed attributed to the magnitude of the drifts with respect to the target model.
\begin{figure}[t]
    \centering
    \includegraphics[width=0.45\textwidth]{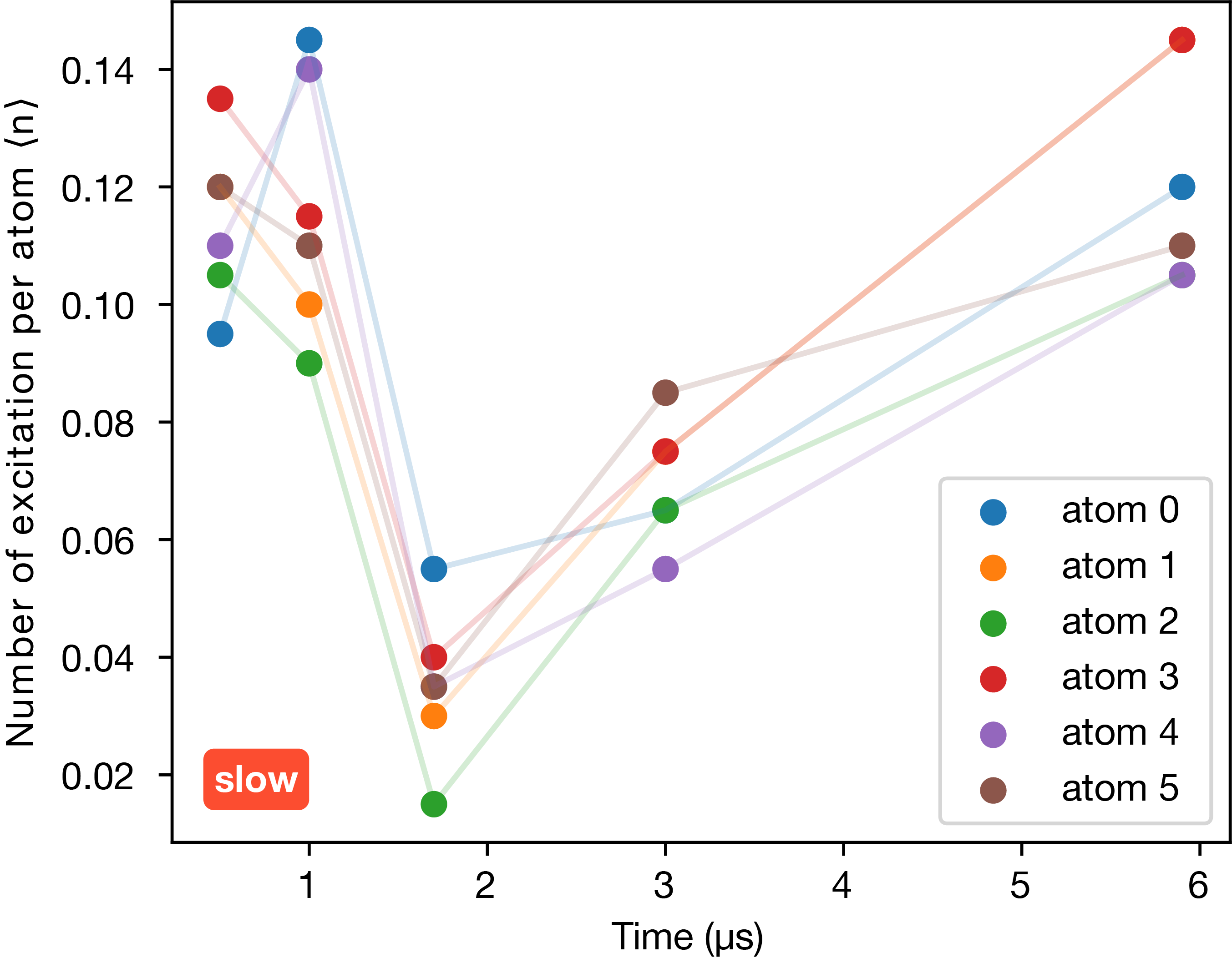}
    \caption{\textbf{Individual atom excitation of slow sequence on Fresnel.} Time-evolution of the excitation of each individual atom for the slow sequence on Pasqal's Fresnel machine. No atom is definitively lost from the traps during the sequence.}
    \label{fig:individual_atoms}
\end{figure}

\bibliography{main.bbl}
\end{document}